# Renal digital pathology visual knowledge search platform based on language large model and book knowledge

Xiaomin Lv[1], Chong Lai[2], Liya Ding[3], Maode Lai[4,*] and Qingrong Sun[1,*]

[1]College of Information Technology, ZheJiang Shuren University, Hangzhou, 310018, Hangzhou, China.

[2]Department of Urology, The First Affiliated Hospital, School of Medicine, Zhejiang University, Hangzhou, 310003, China

[3]Department of Pathology, Research Unit of Intelligence Classification of Tumor Pathology and Precision Therapy, Chinese Academy of Medical Science, Alibaba-Zhejiang University Joint Research Center of Future Digital Healthcare, Key Laboratory of Disease Proteomics of Zhejiang Province, Zhejiang University School of Medicine, Hangzhou 310058, China

[4]Department of Pathology, Research Unit of Intelligence Classification of Tumor Pathology and Precision Therapy, Chinese Academy of Medical Science, Alibaba-Zhejiang University Joint Research Center of Future Digital Healthcare, Key Laboratory of Disease Proteomics of Zhejiang Province, Zhejiang University School of Medicine, Hangzhou 310058, China

*Correspondence: lmd@zju.edu.cn (Maode Lai) and sunqingrong@zju.edu.cn (Qingrong Sun)

**Abstract**

Large models have become mainstream, yet their applications in digital pathology still require exploration. Meanwhile renal pathology images play an important role in the diagnosis of renal diseases. We conducted image segmentation and paired corresponding text descriptions based on 60 books for renal pathology, clustering analysis for all image and text description features based on large models, ultimately building a retrieval system based on the semantic features of large models. Based above analysis, we established a knowledge base of 10,317 renal pathology images and paired corresponding text descriptions, and then we evaluated the semantic feature capabilities of 4 large models, including GPT2, gemma, LLma and Qwen, and the image-based feature capabilities of dinov2 large model. Furthermore, we built a semantic retrieval system to retrieve pathological images based on text descriptions, and named RppD (aidp.zjsru.edu.cn).

## Introduction

Histopathology holds a preeminent position within the diagnostic framework of a multitude of renal afflictions[1], including Acute kidney injury[2] to chronic glomerular inflammation[3, 4], renal organ transplantation[5], and renal malignancies[6] etc. Given the pivotal role that histopathological analysis plays in informing therapeutic strategies and prognostic assessments, seasoned investigators and clinicians have devoted substantial efforts to compile exhaustive book of prototypical histological samples, documenting the hallmark histopathological hallmarks distinctive to each disease phenotype. While numerous books provide a wealth of cases for study and research, readers often lack the capability to promptly retrieve relevant images for real-time clinical cases to give a precise diagnosis in practical diagnostic process.

The advent of large language models has revolutionized the rapid construction and retrieval of knowledge bases, offering a more efficient approach. Semantic generation models such as GPT2 (OpenAI), LLma (Facebook), gemma (Google), and Qwen (Alibaba) exhibit remarkable performance in extracting semantic features from text. Concurrently, knowledge base construction platforms based on large language models, such as OLLMA (https://ollama.com/) and chatRTX (https://www.nvidia.com/en-us/ai-on-rtx/chatrtx/), have emerged. These tools excel in summarizing and retrieving information from extensive model file knowledge bases. In addition, there are also many studies devoted to building foundation models based on massive digital pathological image data from the network. Such as a foundation model for 1.17 million image–caption pairs through CONtrastive learning from Captions for Histopathology (CONCH)[7], a general-purpose self-supervised model for pathology, pretrained using more than 100 million images from over 100,000 diagnostic H&E-stained WSIs (> 77 TB of data) across 20 major tissue types[8] and MONET model based on 105,550 dermatological images paired with natural language descriptions from a large collection of medical literature[9]. However, they fall short in effectively associating text with images within the files.

To address this shortfall, we have undertaken a systematic effort to collate renal histopathology books spanning both Chinese and English sources. This study involved manually segmenting the images and pairing them with their corresponding descriptive narratives. Through this process, subjected to semantic feature extraction using LLma, Gemma, GPT-2, and Qwen large models, which we then constructed a comprehensive renal pathology knowledge base. Subsequently, the resulting sets of features extracted by each model underwent cluster analysis, thereby enabling the identification of distinct distribution and relationships among the semantic content. This analytical step culminated in the establishment of a semantic search system based on flask tool from python, grounded in the textual descriptions of the renal pathology knowledge base. This system enables users to query the repository based on the underlying meaning of their queries, rather than relying solely on keyword matching, thereby facilitating more nuanced and contextually relevant exploration of the rich visual and textual content within the knowledge base.

## Methods

We have curated books containing renal pathology information from PDF Room (https://pdfroom.com/), Pathology Outlines (https://www.pathologyoutlines.com/), Taobao

(https://www.taobao.com/), amazon (https://www.amazon.com/), SPRINGER NATURE (https://www.springernature.com/gp/products/books) 和 China Academic Digital Associative Library (CADAL, https://cadal.edu.cn/), split them into individual pages, segmented the image content to construct a dataset based on image-text pairs. Utilizing large models for semantic encoding and cluster analysis, we ultimately developed a semantic retrieval system based on Flask tool for efficient information retrieval (Figure 1).

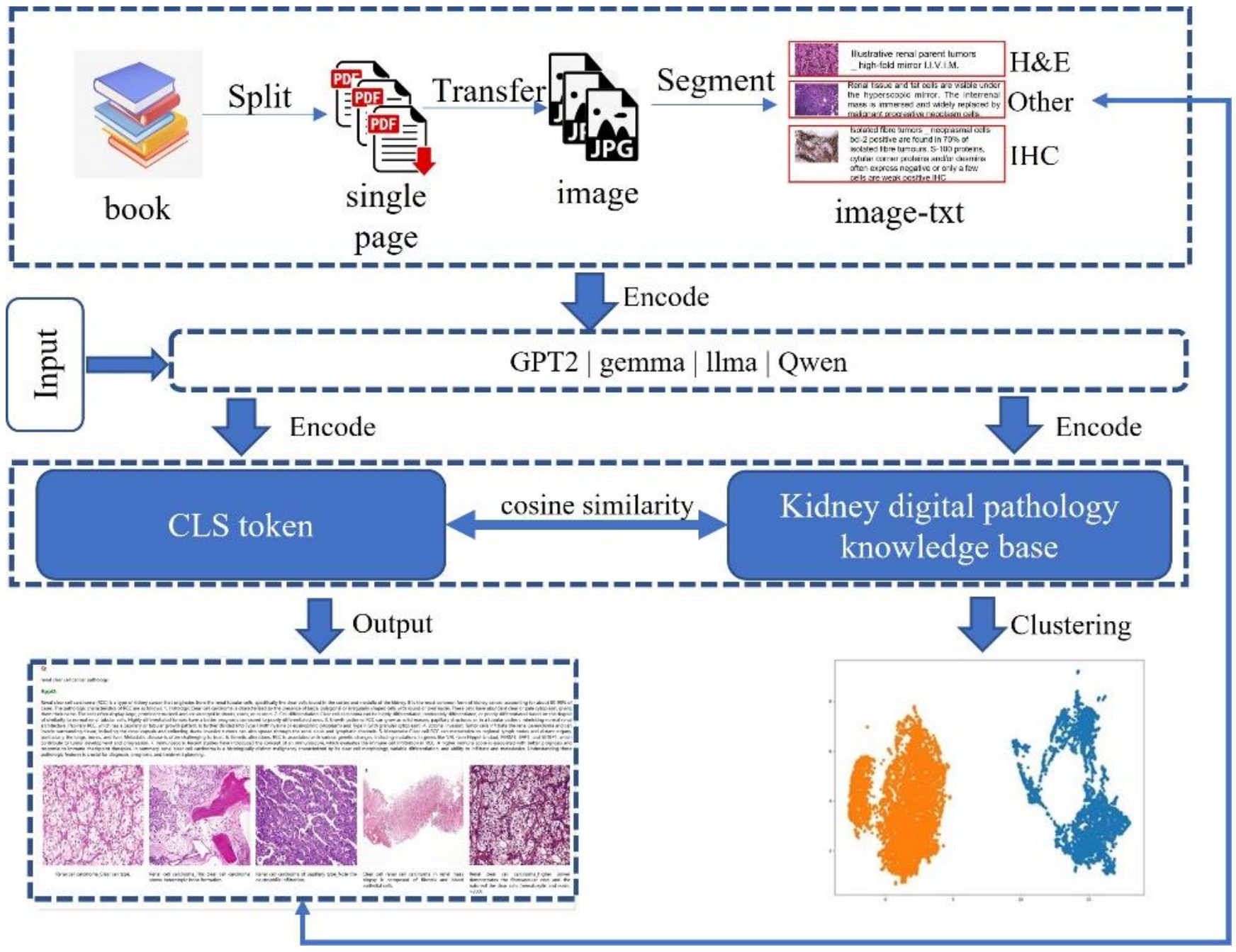

Figure 1. Framework for all progress. H&E means hematoxylin-eosin staining, IHC means Immunohistochemistry, CLS token means Classification token.

1. Image annotation

We utilized the PyPDF2 tool to split the sections containing renal histopathology in all books into individual pages. Subsequently, we used the pdf2image tool to convert each page from PDF to JPG images. Then, employing a labeling tool, we manually segmented the pathological images and paired them with corresponding text descriptions. Ultimately, we obtained a dataset of image-text pairs, for which we constructed index IDs. Each ID consists of nine digits, starting with 'P' followed by 8 valid numbers. The data was categorized into Chinese and English based on language and into immunohistochemistry and pathological images based on image type.

2. Semantic encoding based on large models

We used large models such as GPT2, LLma, gemma and Qwen to extract features from our descriptive text, obtaining tensor data from the last hidden layer of each description file. We then averaged this tensor data to derive text feature vectors, which were subsequently used for clustering and retrieval task design. Finally, we constructed a tensor dataset and corresponding file IDs to create an hdf5 file, establishing a knowledge base for renal histopathological descriptions.

3. Cluster analysis for knowledge base

We utilized the UMAP (Uniform Manifold Approximation and Projection) tool in Python to conduct

dimensionality reduction analysis on the semantic features within the knowledge base. Subsequently, based on the results of dimensionality reduction, we performed cluster analysis using the KMeans function from the scikit-learn tool. Finally, we visualized the clustering results using the matplotlib tool.

4. Semantic retrieval system (RppD)

We utilized the Flask tool in Python to build a semantic retrieval system. This system primarily involves inputting text queries, encoding the text using large models, calculating the cosine similarity between semantic features to evaluate similar pathological description information, and providing the top five most similar pathological images based on IDs.

## Results

We selected 60 out of 344 books containing renal pathology (37 in Chinese, 23 in English), finding a total of 10,317 pathological patch image-text pairs. Four knowledge bases for renal histopathological descriptions were constructed based on semantic feature vectors from four large models. From Figure 2, we can observe the following statistics: 9,342 pathological images, 975 immunohistochemistry images; 4,214 from Chinese sources, 6,103 from English sources; 2,038 related to tumor diseases, and 8,279 related to other diseases. Additionally, word clouds were generated for Chinese and English texts to visually highlight the main keywords.

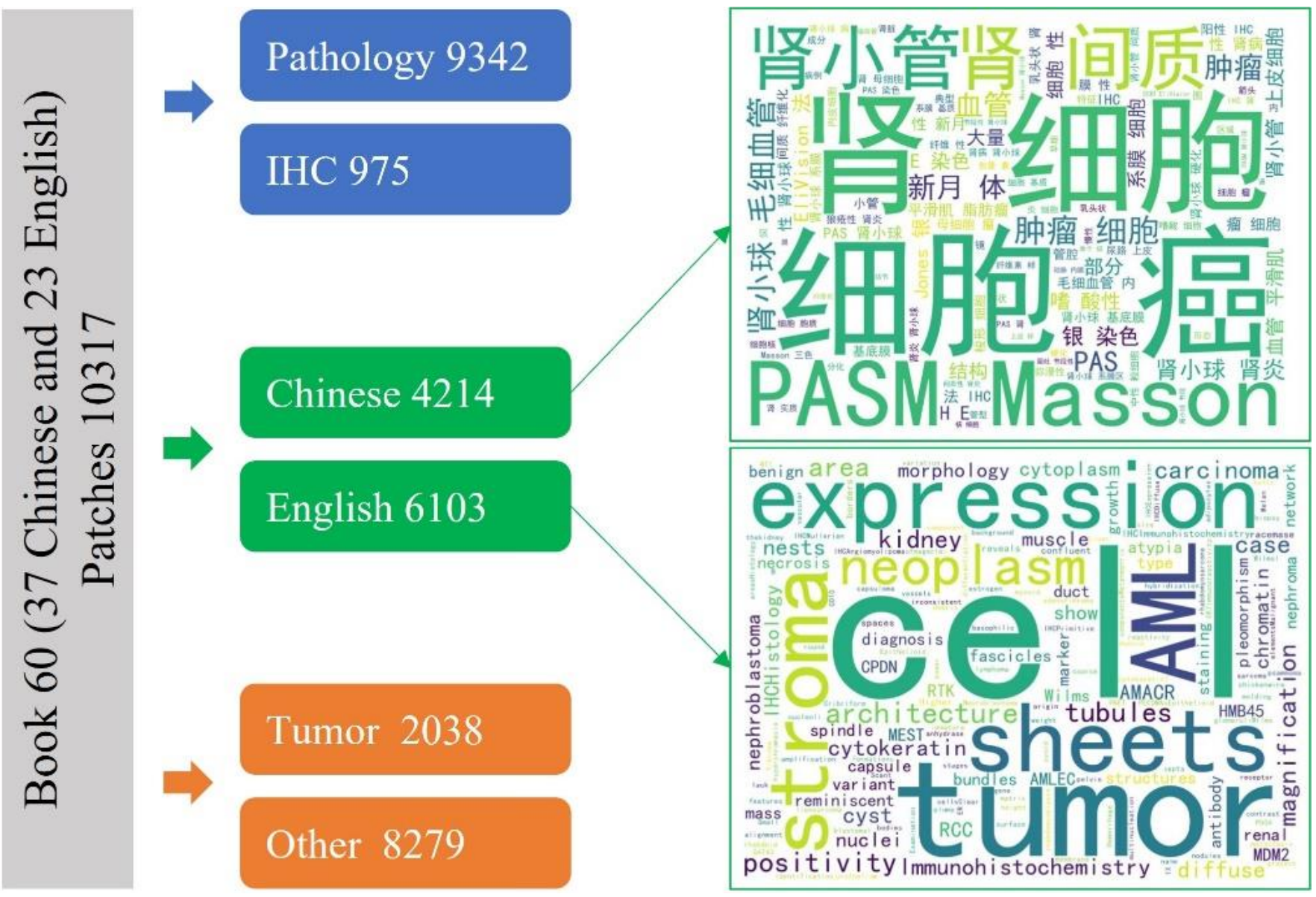

Figure 2. The detail information for renal knowledge base. IHC means Immunohistochemistry.

1. Cluster analysis based on renal knowledge base

Regarding the renal pathology knowledge bases built for the four models, cluster analysis was conducted on all knowledge, revealing that only GPT2 and Qwen effectively separated the Chinese and English content within the knowledge base. Additionally, it was observed that there are certain differences in the semantic features among the various large models (Figure 3).

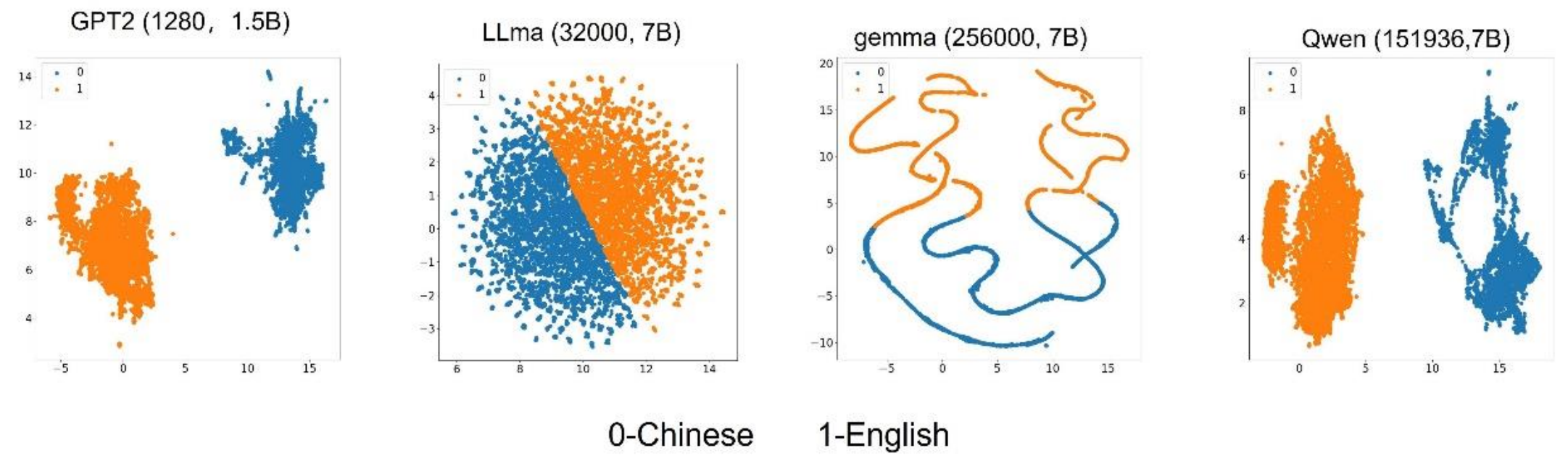

Figure 3. cluster analysis for 4 renal knowledge bases based 4 large models.

To further analyze the distribution and differences of semantic features in the large models, we conducted cluster analysis separately on the Chinese and English semantic features within the knowledge base. From Figure 4, it was observed that the Qwen and GPT2 models exhibited distinct localized distributions in both Chinese and English, which correlated with the actual semantic content. The LLma and gemma models did not show significant distribution specificity in either language. Additionally, LLma and gemma demonstrated clear distribution differences in Chinese and English features, possibly due to gemma primarily operating in an English semantic space.

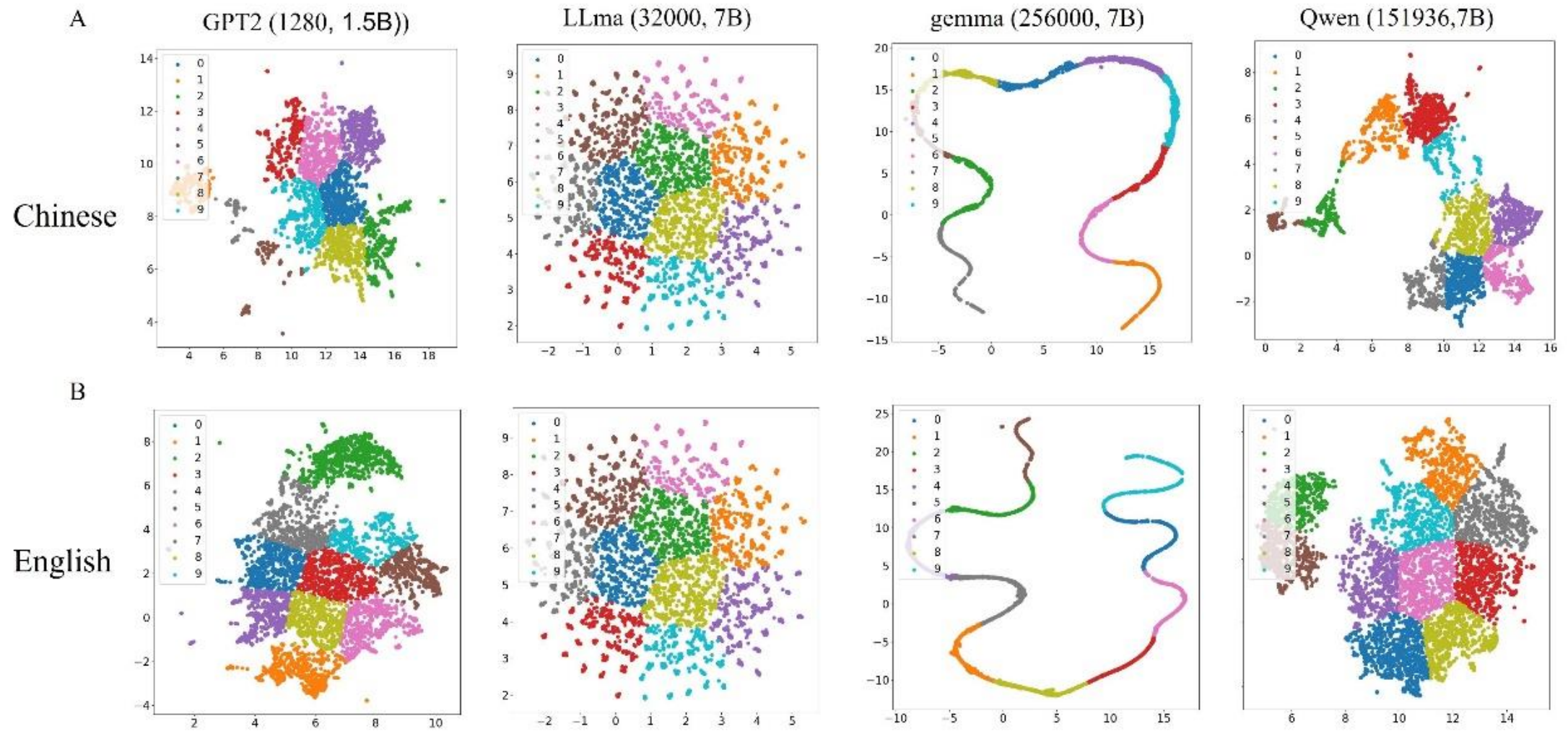

Figure 4. cluster analysis for the Chinese and English semantic features. All cluster analysis results are clustering in 10 classes.

2. Construction of semantic retrieval system

Based on the knowledge bases from the four large models, we developed a semantic retrieval system that supports semantic feature comparison across different models (Figure 5). Additionally, we integrated Qwen's generative mode API into the semantic comparison process. This system first analyzes user queries and then displays the top five matched pathological images. To enhance retrieval accuracy, we set specific threshold values for cosine similarity. If the threshold is not met, the system will return empty results (Figure 5B).

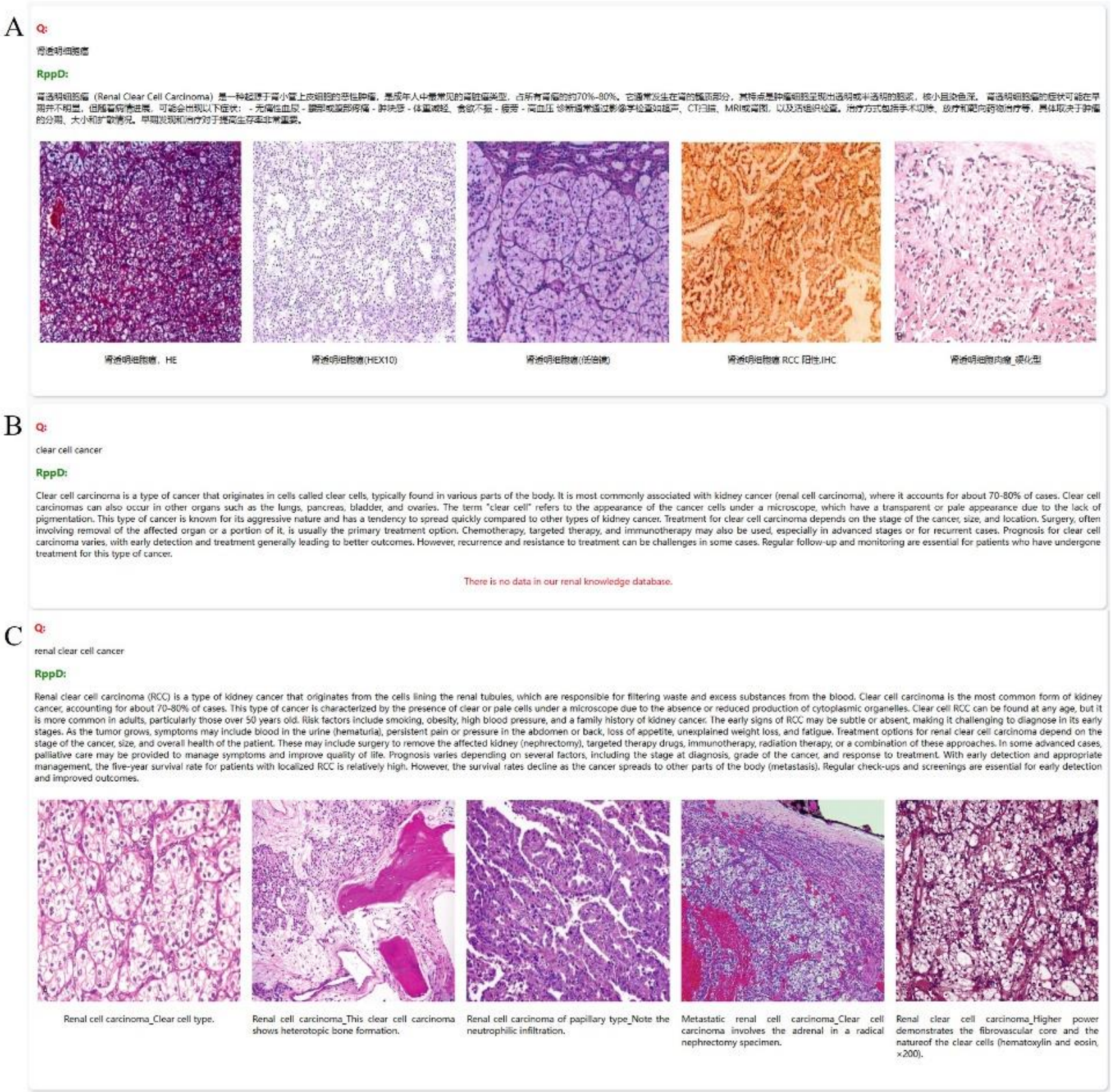

Figure 5. The retrieval results based on semantic retrieval system.

**Discussion**

Through the analysis above, we have established a renal pathology knowledge base containing 10,317 digital pathological images, with detailed categorization based on image type (9,342 pathological images, 975 immunohistochemistry images), language (4,214 from Chinese sources, 6,103 from English sources), and disease types (2,038 related to tumor diseases, 8,279 related to other diseases). Subsequently, we developed a semantic retrieval system based on this knowledge base, accessible at the following website: aidp.zjsru.edu.cn.

The differences in the resources trained by the large models indicate significant variations in their ability to distinguish between Chinese and English. Figure 3 shows that the semantic features extracted by the GPT2 and Qwen models align closely with the actual semantics of our data. This suggests that both models have been trained on medical content, resulting in semantic features that closely resemble medical reality. The differences among the four large models highlight the importance of combining domain-specific knowledge with pre-trained models for extracting domain-specific semantic features.

Furthermore, the extraction of pathological image features also relies on preliminary training data. We conducted cluster analysis based on image features using the dinov2 (from facebook) large models, but the clustering results did not reflect any meaningful sub-class patterns, including differential diseases and stains (as shown in Figure 6).

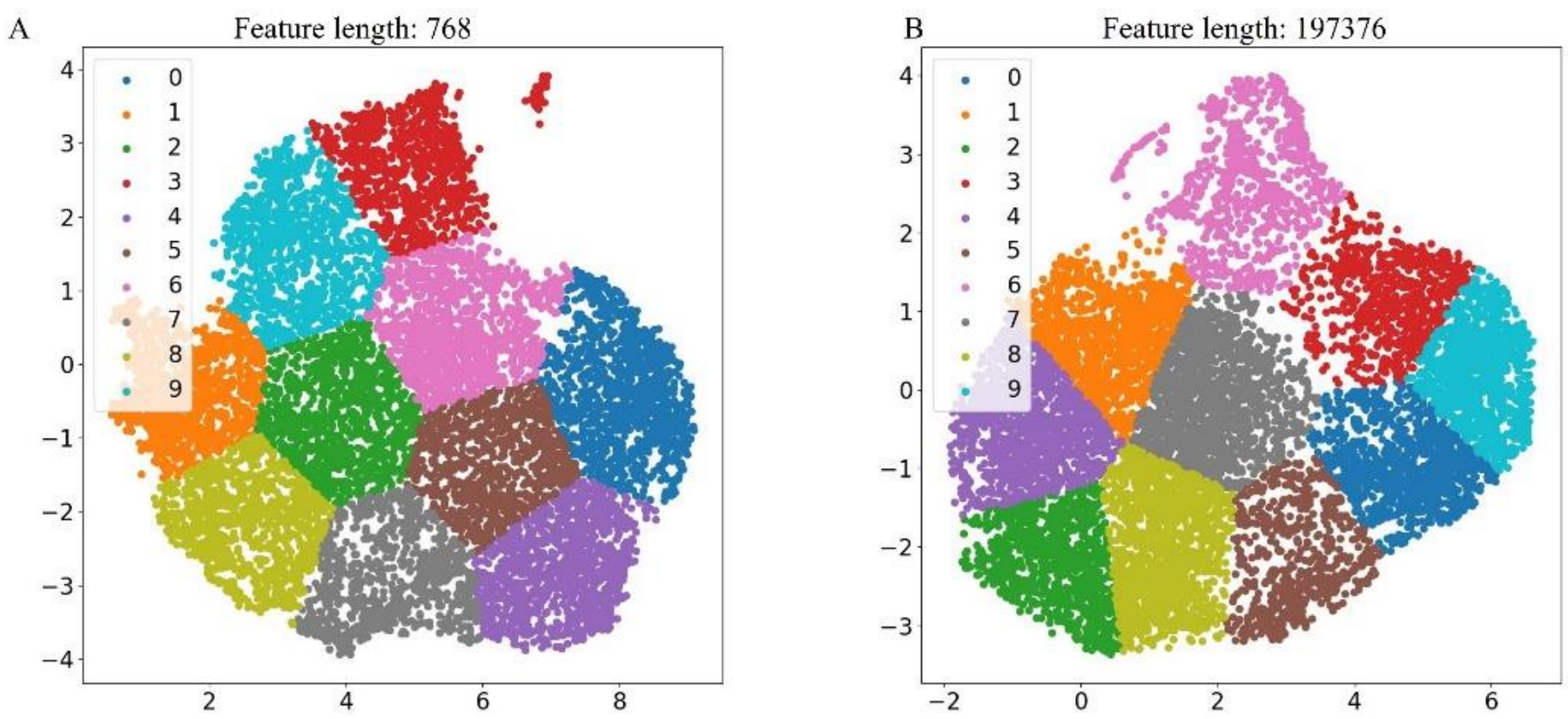

Figure 6. Cluster analysis of pathological image feature based on dinov2 large model. A show the cluster results by mean (last_hidden_states, dim=1); B shows the cluster results based on last_hidden_states.view(-1).

While large models have made significant advancements in general-purpose generation tasks, their applicability in specialized domains still requires additional training data. Nevertheless, several outstanding domain-specific models, such as content generation, sentiment analysis, translation and localization question answering, search and recommendation, education and code developments etc., have achieved remarkable success. Based on these differences, we believe that domain-specific features will be a key challenge in the integration of large models.

## Author Contributions

Q.S. and M.L. conceived the original idea and designed the study. L.D collected book data. X.L, Q.S, C.L and L.D contributed to annotate all pathology images, text description and build dataset. Q.S. and M.L. reviewed and contributed to the manuscript.

## Code Availability

The code resource was released in aidp.zjsru.edu.cn .

## Acknowledgments

This work was supported by the National Natural Science Foundation of China (No. 82072811).

## Competing Interests

The authors declare that they have no competing interests.